# The Importance of Prioritizing Exoplanet Experimental Facilities

A White Paper for the Planetary Science and Astrobiology Decadal Survey 2023-2032

**Erika Kohler** *(301-614-5756, Erika.kohler@nasa.gov , NASA Goddard Space Flight Center (GSFC), Greenbelt, MD, USA)*

**Co-Authors:** Chao He (Johns Hopkins University, Baltimore, MD); Sarah E. Moran (Johns Hopkins University, Baltimore, MD); S.-H. Dan Shim (Arizona State University, Tempe, AZ); Karalee K. Brugman (Arizona State University, Tempe, AZ*); Aleisha C. Johnson (Arizona State University, Tempe, AZ**); Pilar C. Vergeli (Arizona State University, Tempe, AZ); Maggie A. Thompson (University of California, Santa Cruz, CA); Heather Graham (NASA GSFC/CRESST, Greenbelt, MD); Murthy S. Gudipati (JPL, Caltech, CA); Benjamin Fleury (JPL, Caltech, CA); Bryana L. Henderson (JPL, Caltech, CA)

* Fall 2020: Carnegie Institution for Science, Washington, D.C.
** Fall 2020: University of Chicago, IL

**Co-Signers:** Jonathan Fortney (University of California, Santa Cruz); Chuanfei Dong (Princeton University); Christy B. Till (Arizona State University); Monica Vidaurri (Howard University/NASA GSFC); Ravi Kopparapu (NASA GSFC); Noam Izenberg (JHU/APL); Thomas Fauchez (NASA GSFC / USRA); Ehsan Gharib-Nezhad (NASA ARC); Giada Arney (NASA GSFC); Elisa Quintana (NASA GSFC); Lynnae C. Quick (NASA GSFC); Shiblee Barua (NASA GSFC/USRA); M. J. Way (NASA/GISS); Victoria Meadows (University of Washington); Edward Schwieterman (University of California, Riverside); Sarah Horst (Johns Hopkins University); Peter Gao (UC Berkeley and UC Santa Cruz); Joe P. Renaud (USRA/NASA Goddard)

**Introduction:**

A decade ago the exoplanet community was anticipative of the types of planets expected to be discovered from observations from the Kepler mission, TESS (Transiting Exoplanet Survey Satellite), and ground-based telescopes. Now, we anxiously await the observations from RST (Nancy Grace Roman Space Telescope), JWST (James Webb Space Telescope), and future space-based mission concepts like LUVOIR (Large Ultraviolet Optical Infrared Surveyor), HabEx (Habitable Exoplanet Imaging Mission), and OST (Origins Space Telescope), which will allow us to characterize these newly discovered worlds and uncover the secrets they hold. The continuous improvement of instruments and telescopes provide us with a wealth of data used to quantify and characterize the environments, and evolution, of exoplanets. These observations, combined with theoretical modeling, help us to understand the physical and chemical processes shaping exoplanets. However, while the improvements in observational capability will help us address many outstanding questions in exoplanet science, the observations themselves can only be understood in concert with supporting theoretical and laboratory research. By addressing some of the gaps in our observational data, theoretical models will face new tests. But for these tests to be conclusive, the theoretical models need to improve through the inclusion of input data on core chemical and physical properties and processes, for which laboratory experimental work is critical.

A wealth of papers have been written detailing the importance of laboratory experiments prioritizing planetary science and astrophysical needs. Ranging from scientific fundamentals like equations of state, to more precise needs like a spectral database of organic hazes, the main consensus is that laboratory measurements at non-solar system regimes are desperately needed. At present, inputs for models of exoplanet atmospheres depend heavily on extrapolations from terrestrial conditions and assumptions unverified by laboratory studies. Such extrapolations are of dubious value for understanding the atmospheres of exoplanets; in fact, uncertainties in model inputs only serve to produce uncertainties in the model outputs.

These uncertainties extend to exoplanet observations as well. Interpretations of mission and ground-based telescope observations require an accurate understanding of the chemical and physical properties of planetary materials, regardless if the target is an exoplanet or one from our local neighborhood. Laboratory experiments simulating the extreme conditions found on other planets will only serve to maximize the scientific return of observations, from missions and ground telescopes alike. As instruments and techniques advance laboratory data will become even more important. Our greatest chance of success in interpreting and understanding exoplanets will require a three-pronged approach involving observations, modelling efforts, and laboratory experiments.

*In situ* and analytical experiments are necessary for the interpretation of exoplanet observations, and the verification of theoretical models. It is critical that laboratory work receives the funding and support it needs to develop and maintain the facilities necessary to provide critically needed chemical, and physical data. While laboratory studies are often deferred due to their high expenses, with the discovery of exotic regimes on exoplanets, the data generated by these studies are needed now more than ever. The rest of this paper will highlight some of the experimental facilities that currently exist, along with their accomplishments, and a discussion of future facilities that are necessary for moving our knowledge of exoplanets forward in the next decade.

**Experimental development:**

For the purposes of this paper, an "experimental facility" is a chamber or reaction vessel, along with its heating/cooling system, gas handling system, and data collection system. The primary objective of these facilities is to simulate the environments observed on exoplanets while collecting data through some type of measurement. Ideally, these systems are used to vary the chemistry, temperature, and pressure in the vessel, while often, like in the case of spectroscopy, taking measurements over a range of wavelengths and using other analytical tools such as mass spectrometry to obtain deeper insight into the processes. Typically each facility is developed to answer one science investigation, and can sometimes evolve to be used for other applications as they arise. It is important to note that as each science investigation is varied, the resulting experimental facilities are just as varied. For example, it is unlikely that one facility designed to take one type of measurement would be capable of simulating every conceivable temperature and/or pressure. As technology advances so too will our experimental capabilities. Hence, complementary facilities should be developed across the USA that provide complementary data for a comprehensive understanding of processes involved.

Often laboratory work is passed over because it is frequently viewed as cost prohibitive. Laboratory research requires equipment purchases and extensive labor costs. Each experimental facility is typically the result of hours to years involving design, technology development, invention, and methodology development -- before the facility is even built. It takes on an average 3-5 years for a laboratory facility to be developed, tested, and calibrated. Once built, the facility needs to continue to be maintained in addition to the time spent taking the measurements themselves. Long-term sustainable funding is critical to leverage best science out of these facilities. Equipment and laboratory testing of exotic regimes is expensive and time consuming. However, the scientific return in taking *in situ* and analytical measurements in these unusual environments is high. These measurements are absolutely necessary to maximize our understanding of exoplanets.

**Current facilities and their major accomplishments:**

Despite the difficulties faced in developing exoplanet laboratory facilities, there are several labs that are providing valuable research and advancing our understanding of exoplanets. Table 1 lists the facilities that are currently contributing to exoplanet research.

**Table 1.** Summary of available exoplanet facilities.

| Facility | Location | Temperature range (°C) | Pressure range (bar) | Input samples | Notes |
|---|---|---|---|---|---|
| TGA | NASA GSFC | Ambient – 1700 | High vacuum ($\sim 10^{-10}$) – ambient | Solid minerals | Measures mass loss as function of temperature to determine vapor pressures of refractory materials |
| Aabspec cell | NASA GSFC | -170 – 950 | High vacuum – 133 | $CO_2$, $N_2$, $SO_2$ | 6" x 3"; High temperature and variable pressure cell for IR spectroscopy |
| PHAZER | JHU | -196 – 527 | $10^{-4} - 10^{-2}$ | $H_2$, He, $N_2$, $H_2O$, CO, $CO_2$, $NH_3$, $CH_4$, $H_2S$, $SO_2$ | Generation of haze analogues with both plasma and UV source. RGA for gas product identification. FTIR for spectral characterization |
| Furnace & RGA | UC Santa Cruz | $\leq 1200$ | High vacuum ($\sim 10^{-8}$) | Meteorite samples | Monitor outgassing abundances as a function of temperature of volatile species from meteorite samples to inform exoplanet atmospheres that form via outgassing |
| End-loaded piston-cylinder | Arizona State University | $\leq 1700$ | 8000 – 35,000 | Silicates, oxides, metals, volatiles; solids and liquids | Generation of pressure-temperature-composition for melts (magma) produced at depths corresponding to the shallow mantle to crust (epic.asu.edu) |
| Diamond anvil cell | Arizona State University | 25–6000 | 1 – 3e6 | $H_2$, He, $N_2$, $H_2O$, $CO_2$, $NH_3$, $CH_4$ Silicates, oxides, metals; liquid or solid | Generation of pressure-temperature conditions expected for interiors of a range of planets (rocky, ice giants, and gas giants) https://www.danshimlab.info |
| COSmIC | NASA ARC | -223 to ambient | $10^{-4} - 10^{-2}$ | $N_2$, $CH_4$ (so far) | https://www.nasa.gov/content/cosmic-lab; various spectrometers |
| CAAPSE | JPL | Ambient – 1500 | $10^{-4} - 10^{-1}$ | $H_2$, $H_2O$, CO, and their isotopes | e.g. Fleury et al., 2019; UV discharge source, FTIR, QMS |

For planetary scientists much of our focus has been concentrated on exoplanet atmospheres, largely due to the types of exoplanet observations that currently exist. These observations thus far primarily provide atmospheric spectroscopy of larger gaseous planets. As a result, most of the exoplanet facilities aim to provide information related to atmospheric chemistry, clouds, and hazes. The PHAZER (Planetary HAZE Research) chamber at Johns Hopkins University (JHU) simulates haze formation in a variety of exoplanet atmospheres to study photochemical processes in exoplanetary atmospheres. The PHAZER chamber allows investigations with two different energy sources (cold plasma and UV photons) that simulate different energetic processes in planetary atmospheres. After completion of the experiments, the samples are collected (and if necessary stored) in a dry ($H_2O$, $O_2$ free) $N_2$ glove box until subsequent analysis to minimize the possibility of contamination through interactions with Earth's atmosphere. COSmIC (Cosmic Simulation Chamber) at NASA Ames (ARC) also simulates and characterizes haze formation in cooler atmospheres. Outgassing experiments on chondritic meteorites are being performed at UC Santa Cruz in a furnace retrofitted with a Residual Gas Analyzer (RGA) mass spectrometer. Results from this facility provide partial pressures, mole fractions and relative abundances of outgassed volatile species (e.g., $H_2O$, $CO$, $CO_2$, $H_2$, $H_2S$, $CH_4$) from each meteorite sample as a function of temperature to which the samples are heated. This facility provides laboratory data to compare with theoretical modeling of expected compositions for early terrestrial exoplanet atmospheres that form via outgassing.

High temperature photochemical processes and haze formation in a variety of exoplanet atmospheres and surfaces are investigated with the CAAPSE (Cell for Atmospheric and Aerosol Photochemistry Simulations of Exoplanets) facility at the Jet Propulsion Laboratory (JPL) using UV discharge lamps coupled with a high-temperature furnace tube (rated to 1500 °C) at pressures below 1 bar. Mass spectrometry and IR spectroscopy techniques are used to investigate how thermal and photochemical processes affect gas reactivity and the formation of aerosols. One key limitation for the development of laboratory simulations is the possibility to reproduce extreme conditions (high temperature, UV fluxes, etc.) in a laboratory setting. Higher temperature gas cell furnaces, reaching up to 2500°C, and UV transparent (<100 nm) vacuum optics would be desirable, but these temperatures are currently limited by both the availability of appropriate gas cell and optics materials and the availability of tube furnaces in this range. Custom development of these components and furnace systems would be needed to enable coverage of the full temperature range of hot exoplanet atmospheres and reproduce the full UV wavelength range for initiating reactions in exoplanetary atmospheres.

NASA Goddard Space Flight Center (GSFC) hosts a suite of exoplanet facilities designed to simulate various aspects of exoplanetary atmospheres. The thermogravimetric (TGA) system has measured the vapor pressure of several proposed atmospheric constituents of hot Jupiters aimed to ensure that accurate thermodynamic inputs are used for exoplanet cloud formation models. A high temperature/low pressure chamber, capable of reaching 2400°C, is being built which will allow for the same mass loss measurements but at higher temperatures relevant to the atmospheres of brown dwarfs. This chamber has several external ports which allow for *in situ* measurements from a variety of instruments. To ensure maximum scientific return from spectroscopic observations, Goddard takes *in situ* near- to mid-IR measurements with a Fourier Transform InfraRed (FTIR) spectrometer retrofitted with an AABSPEC #2000-A multimode system that houses samples and allows them to be heated up to temperatures of 950°C under pressures of 133 bar–$10^{-10}$ bar. The results of these experiments indicate a diversity of spectrographic features for atmospheric particles that change with varying temperatures. These experiments demonstrate the importance of having laboratory data that are used to calibrate and understand exoplanet atmospheric observations.

Our interest in exoplanets is not limited to only the atmosphere, but also extends to the surface and below. At Arizona State University (ASU), the Metals, Environmental and Terrestrial Analytical Laboratory (METAL) hosts an anaerobic chamber, low $O_2$ optical sensors, a custom quartz reactor, and a 1000 W Xe arc lamp solar simulator with accompanying optical filters to measure the nature and kinetics of aqueous chemical reactions under controlled $pO_2$, $pCO_2$, and UV flux. The chemical constraints identified by such experiments are relevant to understanding weathering as well as photo-reductive/oxidative processes that likely proceed on the surface of rocky exoplanets.

Scientists at ASU's Experimental Petrology and Igneous processes Center (EPIC) are running end-loaded piston-cylinder experiments to predict the composition of magmas that may be produced by partial melting of hypothetical exoplanet silicate mantles, as well as the location of the solidus for these exoplanet compositions. As these mantle-derived melts contribute to the composition of a planet's crust, this is a first step in understanding the compositions of exoplanet surfaces that will be available for interaction via biological and atmospheric chemistry, as well as to build more robust geochemical models for mantle melting in silicate planets that do not have the same composition as Earth.

The diamond-anvil cell lab at ASU can generate pressure and temperature conditions expected for the interiors of planets. The lab is equipped with sample loading systems capable of handling a wide range of materials for exoplanet research, including silicates, oxides, metals, ices, and gases (hydrogen and helium). Combined with synchrotron X-ray diffraction, laser-heated diamond-anvil cells can provide data on equations of state for a wide range of materials at high pressures and high temperatures. The data then can be used for improving mass-radius relations for exoplanet characterization. Diamond-anvil cells enable researchers to understand phase relations and chemical reactions between planet building materials at high pressures and high temperatures relevant for planetary interiors. Such data allow the researchers to link internal processes with the evolution of surface and atmospheres of exoplanets.

Facilities designed for our solar system can also be useful for simulating and understanding exoplanets. Especially in terms of habitability, Earth- and Venus-analogs are of great interest to the community. Venus experimental facilities that focus on scientific investigations for surfaces and atmospheres could be utilized by the exoplanet community to run a range of simulations for a variety of rocky-like planets. AVEC (APL Venus Environment Chamber) is capable of simulating Venus conditions related to temperature, pressure, and chemistry. GSFC manages two Venus simulation chambers of differing sizes that have been used to investigate surface-atmospheric mineralogical interactions and can easily be adjusted to simulate the conditions of cooler (<500°C) exoplanets. A more thorough list on the capabilities of Venus experimental facilities can be found in the White Paper by Santos et al.

**Future facilities:**

Fortney et al. in their 2016 white paper, "The Need for Laboratory Work to Aid in the Understanding of Exoplanetary Atmospheres" state "Uncertainties in a path towards [exoplanet] model advancement stem from insufficiencies in the laboratory data that serve as critical inputs to atmospheric physical and chemical tools." The authors, consisting of some of the world leaders in exoplanet atmosphere modeling, identified several areas where laboratory data is critical for exoplanet research. Some of these needs include molecular opacity linelists with parameters for a variety of broadening gases, high spectral resolution opacity data for a variety of molecular species, and gas photoabsorption cross sections at high temperatures.

These areas will help inform the development of exoplanet experimental facilities. While some exoplanet laboratory studies are nonexistent due to lack of funding, or because exoplanets are still a young field, some require further technological advancement before the experimental facilities can be built. Many of the core chemical, optical, and physical properties, and processes that are needed as model inputs require measurements at physical and chemical domains currently physically impossible to replicate in the laboratory. However, the need for this data is great and will drive actions to develop the technologies necessary to take these measurements.

To answer astrobiological questions beyond the broad categorization of "habitability", an entirely new kind of facility will need to be developed. Simulations that incorporate organism growth could be used to test metabolisms in the context of their geochemistry and provide atmospheric and geologic indications that could be linked to remote observations. While it is assumed that certain gases in particular combinations and ratios can be used as biosignatures, there is little experimental evidence to inform the complex web of feedbacks between biological communities and non-analog exotic lithospheres and hydrospheres to indicate the probability of these gases' metabolic byproducts coexisting in an atmosphere at a time scale that would be remotely observed. Further, there is little research being performed that can assign metabolism networks or community structure from remote observation. In addition to specialized facilities, exoplanet astrobiology requires collaboration with and recruitment of biologists and physiologists to provide credible experimentation on these problems of scale.

**Conclusions and recommendations:**

Approaching exoplanetary research from a big picture perspective, involving a collaborative effort between observers, modelers, and laboratory scientists, will ensure steady advances in the field and will lead to maximum scientific return from current and future exoplanet observations. While funding for laboratory research often gets deferred, or even lags behind mission selections, we have the opportunity to get ahead of the game and encourage unified momentum while exoplanetary research is still young. With this in mind we recommend the following:

1. To prioritize the development and performance of laboratory studies simulating exoplanetary environments.
2. To prioritize the development of long lasting-modular facilities. The cost benefit and scientific payoff is historically high for facilities that can evolve as our science questions evolve when we learn more about these planets.
3. To include laboratory funding in mission proposals and selections; which will substantially improve our capabilities in interpreting exoplanet observations.

The above recommendations will ensure our laboratory facilities are maintained and expanded on, allowing for a unified approach to the field of exoplanetary science. The exoplanet community has increasingly realized the importance of collaborative efforts across scientific disciplines (astrophysics, planetary science, heliophysics, etc.) and scientific methods (observations, models, experiments). From a scientific perspective, funding support of experimental facilities will allow for better interpretations of observations, and an increase in our characterization efforts - thus understanding - of these exotic worlds.